# Study of the radiological doses and hazard indices in soil samples from Karbala city, Iraq

## Mohammed Abdulhussain Al-Kaabi[1, a] and Ahmed Al-Shimary[1,b]

[1]Department of Physics, College of Science, Kerbala University, Iraq
[a] E-mail: corresponding author m_alkaabi71@yahoo.com
[b] E-mail ah_alshimary@yahoo.com

**Abstract:**
The radiological doses and hazard indices of natural radionuclides $^{238}$U, $^{232}$Th and $^{40}$K in soil samples from Kerbala city were evaluated using gamma-ray spectroscopy system using NaI(Tl) "1.5×2" detector in low-background with 24 hour. The average values of absorbed gamma-ray dose rate, annual effective dose equivalent and annual gonadal dose equivalent were found to be 90.83±2.00 nGy/h, 111.89±2.46 μSv/y and 640.85±15.44 μSv/y respectively. The average values of gamma representative level index and external hazard index resulting from natural radionuclides for all samples in the study area were 1.43±0.031 and 0.53±0.011 respectively. The obtained results in current work were compared with other results for different countries.

## 1. Introduction

Naturally radionuclides come from the atmosphere as a result of radiation from outer space, earth's crust such as rocks mineral ores and soil, its emitted from both natural and human-made radionuclides and surrounds us at every time[1].

The naturally radionuclides materials that have very long half-lives include $^{238}$U, $^{235}$U and $^{232}$Th chains (terrestrial radionuclides) distribute widely on earth and ocean, they were already present when the earth was born about 4.5 billion years ago and each of these nuclides terminates in stable isotopes of (Pb) nuclide[2]. The other naturally radionuclides such as $^{40}$K, $^{87}$Rb, and $^{113}$Cd are individual and the most important ($^{40}$K) a radioactive isotope with a long half-life (1.28 x 10$^9$ years) [3]. It is widely distributed on earth and found in measurable quantities in many building materials [4]. There are also naturally radionuclides originated from interactions between the cosmic ray and the outer atmosphere such as $^3$H, $^{14}$C, $^{14}$N, $^{81}$Kr, $^{22}$Na[2]. Man-made radionuclides, which are present in environment, have been created by human activities and added to the inventory of natural radionuclides for example $^3$H, $^{131}$I, $^{129}$I, $^{137}$Cs, $^{90}$Sr and $^{239}$Pu, in spite of the amount added is little compared to natural quantities. In 1996, IAEA estimated that 20% of doses contribution in the environment are come from cosmic ray and man-made processes while 80% is from the natural radionuclides [5]. The presence of the radionuclides depends on the geological and geographical conditions, therefore we can find different levels of radionuclides in the soil samples in different region in the world [6].

The aim of this study was to estimate the levels of radiological doses as absorbed gamma-ray dose rate, Annual effective dose equivalent and annual gonadal dose equivalent in the soil samples collected from different locations in Kerbala city, Iraq. Moreover, radiological hazard indices as gamma representative level index and external hazard index were calculated and compared with results for different regions in the world.

## 2. Materials and methods

*2.1. Samples collection and preparation*

The thirty soil samples were collected from different sites of Karbala city during October and November 2014. The collected sample taken from random places as in figure 1 with depth of 5 cm. Table 1 show the sampling location at the area under study. The samples were crushed and dried to ensure that any significant moisture was removed. After that a sieve with diameter holes 500 μm was used to obtain a homogeneous powder and then weighed by 1 kg each one. Then the samples were packed into 1 liter polyethylene plastic Marinelli beakers. Plastic Marinelli beakers were sealed with a tape and stored for about one month before counting to allow secular equilibrium to be attained between $^{222}$Rn and its parent $^{226}$Ra in uranium chain [7].

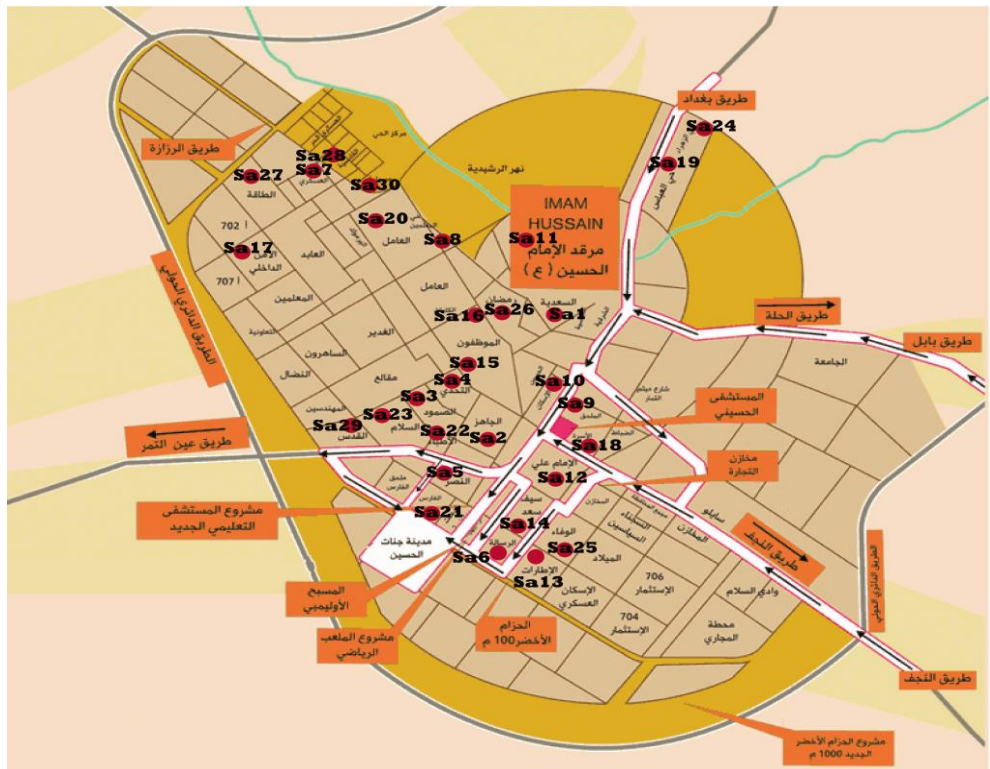

**Figure 1.** Distribution of collective samples on the map of Karbala city.

**Table 1.** Location of the thirty soil samples collected from different districts of Kerbala city.

| Sample No. | Sample Code | Position Longitude | Latitude | Location |
|---|---|---|---|---|
| 1 | Sa1 | 44.015944 | 32.614226 | Jameaa |
| 2 | Sa2 | 44.017038 | 32.587162 | Benaa Jahez |
| 3 | Sa3 | 43.997273 | 32.593584 | Semood |
| 4 | Sa4 | 44.004239 | 32.592088 | Tahady |
| 5 | Sa5 | 44.007504 | 32.576145 | Naser |
| 6 | Sa6 | 44.012931 | 32.560085 | Resala |
| 7 | Sa7 | 43.979555 | 32.614472 | Askaree |
| 8 | Sa8 | 44.001885 | 32.622182 | Moalmeen |
| 9 | Sa9 | 44.022197 | 32.589249 | Askan |
| 10 | Sa10 | 44.041735 | 32.590336 | Molhak |
| 11 | Sa11 | 44.042898 | 32.575158 | Industrial City |
| 12 | Sa12 | 44.0228829 | 32.578023 | Imam Ali |
| 13 | Sa13 | 44.021574 | 32.554645 | Itarat |
| 14 | Sa14 | 44.019797 | 32.572244 | Shohadaa Imam Ali |
| 15 | Sa15 | 44.003453 | 32.601484 | Shohadaa Moadafeen |
| 16 | Sa16 | 44.006756 | 32.603366 | Moadafeen |
| 17 | Sa17 | 43.966443 | 32.618725 | Amen dakhly |
| 18 | Sa18 | 44.032509 | 32.584459 | Osraa |
| 19 | Sa19 | 44.01807 | 32.635158 | Abass |
| 20 | Sa20 | 43.986353 | 32.627845 | Amel |
| 21 | Sa21 | 44.003696 | 32.565916 | Fares |
| 22 | Sa22 | 43.999718 | 32.58.265 | Atebaa |
| 23 | Sa23 | 43.99763 | 32.584237 | Salam |
| 24 | Sa24 | 44.048823 | 32.641497 | Zahraa |
| 25 | Sa25 | 44.026011 | 32.56712 | Wafaa |
| 26 | Sa26 | 44.011411 | 32.604433 | Ramadan |
| 27 | Sa27 | 43.968185 | 32.638977 | Taka |
| 28 | Sa28 | 43.98178 | 32.639969 | Kadesea |
| 29 | Sa29 | 43.987539 | 32.580592 | Qudos |
| 30 | Sa30 | 43.990035 | 32.641174 | Mojtaba |

## 2.2. Gamma-ray spectrometry

Gamma spectrometry system used for the measuring emitted gamma rays from soil samples. The system consist of NaI (TI) scintillation detector, BICRON, crystal dimensions of 1.5×2 inch with CASSY lab program (1024 channel MCA). A detector having 7.5 % energy resolution at 662 keV gamma line of radioactive source $^{137}$Cs. It is surrounded by a shield consist of a cooper of thickness 6 mm and a lead of thickness 4 cm to reduce the background of gamma radiation to minimum. The detector was energy-calibrated using a set of standard gamma-ray radioactive sources are $^{137}$Cs (661.66 keV), $^{22}$N (511, 1274 keV) and $^{60}$Co (1173.24 and 1332.5 keV) to cover a sufficient range of photo peaks. The photo peaks of $^{214}$Bi (609 keV), $^{228}$Ac (338 keV) and $^{40}$K (1460 keV) were used to specifying the activity of $^{238}$U, $^{232}$Th and $^{40}$K in the soil samples.

## 2.3. Background measurement

Before the counting process of the samples, the gamma background was determined with an empty Marinelli beaker for 24 hours in the same manner that used for samples. The background was subtracted from the measured gamma- ray spectra of each sample.

## 3. Results and discussion

### 3.1. Absorbed gamma-ray dose rate (D)

The main contribution to the absorbed dose rate in the air comes from terrestrial gamma-ray radionuclides present in trace amounts in the soil, the measurements of dose rate depend on measurements of specific activity concentrations of radionuclides, mainly $^{238}$U, $^{232}$Th and $^{40}$K families. The 2008 UNSCEAR report explain that the absorbed dose rate $D$ in air 1 meter above the ground surface can be given by [8]:

$$D(nGy/h) = 0.462 A_{^{238}U} + 0.604 A_{^{232}Th} + 0.0417 A_{^{40}K} \qquad (1)$$

Where $A_{^{238}U}$, $A_{^{232}Th}$ and $A_{^{40}K}$ are the specific activities of $^{238}$U, $^{232}$Th and $^{40m}$K families in (Bq/kg) respectively. The dose convention factors of $^{238}$U, $^{232}$Th and $^{40}$K are 0.462, 0.604 and 0.0417 in ((nGy/h)/(Bq/kg)) respectively. Table 2 shows that the absorbed dose rate values for the samples ranged from 58.89±1.79 to 138.25±2.31 nGy/h with an average value of 90.83±2.00 nGy/h. The average values of the absorbed dose rate were calculated for soil sample in some countries as 95.5 nGy/h in India [9], 37.15 nGy/h in Jordan[10], and 64.5±27.1 nGy/h in Thailand[11]. The population-weighted value of the absorbed dose rate in outdoor that calculated by UNSCEAR 2000 was 59 nGy/h. Figure 2 represent the samples and the absorbed dose rate and it is shows also the contributions of $^{238}U, ^{232}Th$ and $^{40}K$ in each sample.

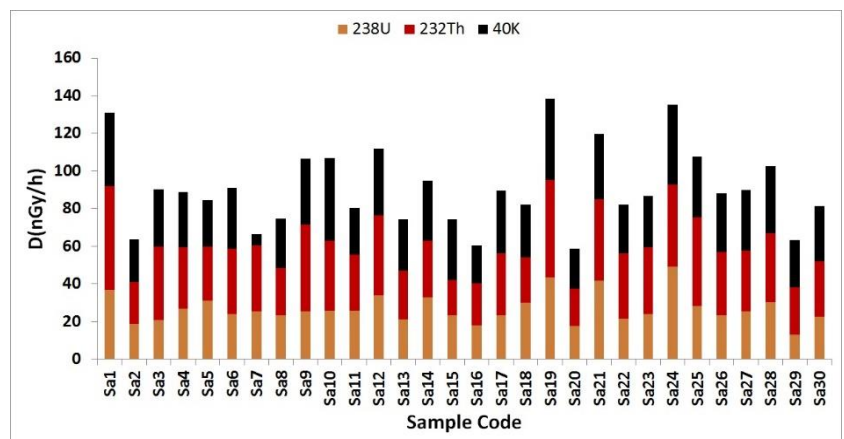

**Figure 2.** The measured absorbed dose rate for all soil sample.

*3.2. Annual effective dose equivalent (AEDE)*

The calculations of effective dose equivalent depend on the value of the absorbed dose rate in air. To accomplish these calculations, account must be taken of the conversion coefficient from absorbed dose rate in air to effective dose equivalent received by adult and occupancy fraction. The value of these two parameters vary depending on the climate at the area considered and the average age of the population. In the UNSCEAR 2008 report, the value of conversion coefficient was 0.7 Sv/Gy for male and female and to the indoor and outdoor, and the 0.2 for the outdoor occupancy fraction. Therefore, the outdoor annual effective dose equivalent can be given as follow [8]:

$$\text{AEDE}(\mu Sv/y) = D(nGy/h) \times 8760(h) \times 0.2 \times 0.7 (Sv/Gy) \times 10^{-3} \quad (2)$$

Table 2 shows the values of the annual effective dose equivalent for the different areas of the soil samples. The values varied from 72.22±2.20 to 169.55±2.83 µSv/y with mean value and standard deviation of 111.89±2.46 µSv/y. The average values of the annual effective dose equivalent were calculated for soil sample in some countries as 314.1 µSv/y in Turkey [12], 32.33 µSv/y in Saudi Arabia [13], 152 µSv/y in China [14]. The worldwide average value calculated by UNSCEAR 2008 was 70 µSv/y. Figure 3 shows the samples and the annual effective dose equivalent.

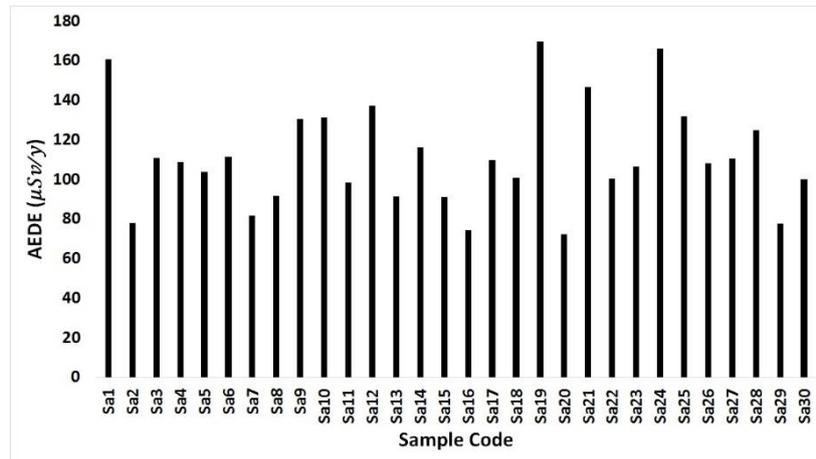

**Figure 3.** The measured annual effective dose equivalent for all soil sample.

*3.3. Annual gonadal dose equivalent (AGDE)*

The organs of interest by UNSCEAR include the thyroid, the lungs, bone marrow, bone surface cell, the gonads and the female breast [8]. Hence, the annual gonadal dose equivalent can be given by [15]:

$$\text{AGDE}(\mu Sv/y) = 3.09\, A_{238_U} + 4.18\, A_{232_{Th}} + 0.314\, A_{40_K} \quad (3)$$

The obtained values of annual gonadal dose equivalent are listed in table 2, the values varied from 416.49±12.70 to 972.97±16.34 µSv/y with mean value and standard deviation of 640.85±15.44 µSv/y. The mean values of soil samples were 439.73 µSv/y for Nigeria [16], 2398 µSv/y for Egypt [17], and 182.52 µSv/y for Saudi Arabia [13]. The contributions of specific activities due to $^{238}U$, $^{232}Th$ and $^{40}K$ in each sample were shown in figure 4.

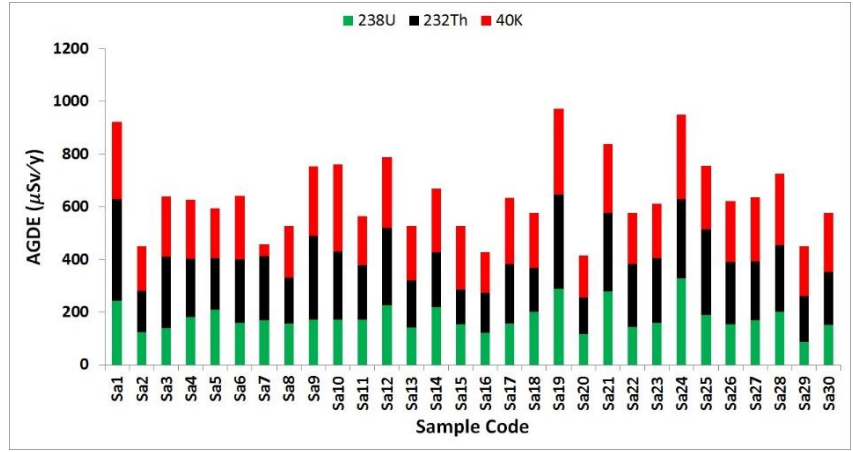

**Figure 4.** The measured annual gonadal dose equivalent for all soil sample.

### 3.4. Gamma representative level index $(I_{\gamma r})$

The gamma radiation representative level index associated with natural radionuclide was evaluated using the following equation [18]

$$I_{\gamma r} = \frac{A_{238_U}}{150} + \frac{A_{232_{Th}}}{100} + \frac{A_{40_K}}{1500} \qquad (4)$$

Table 2 shows the obtained values of gamma index for our samples, the values 0.93±0.03, 2.17±0.04 and 1.43±0.031 indicate a minimum, maximum and mean values for this index. The values for sediments ranged between 0.248 and 2.735 in India [19], 0.89 and 1.03 in Nigeria [16]. The contribution of radioactive chains seems clear from figure 5.

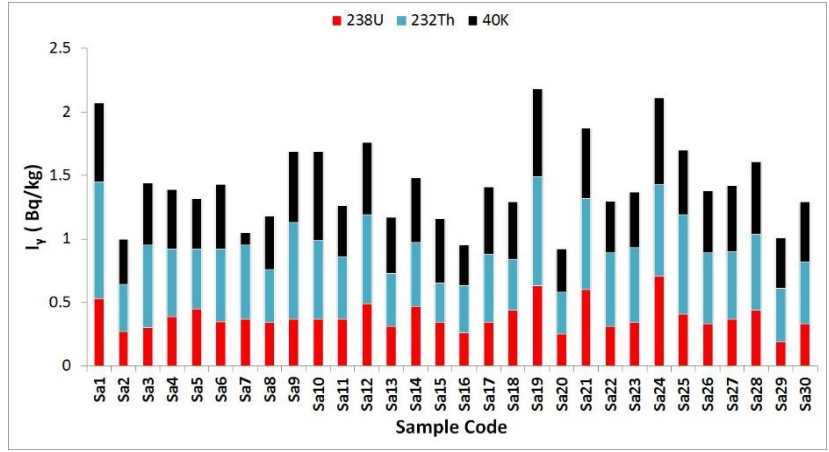

**Figure 5.** The gamma representative level index for all soil sample.

### 3.5. Hazard index $(H_{ex})$

The external hazard index, is defined as [20]

$$H_{ex} = \frac{A_{238_U}}{370} + \frac{A_{232_{Th}}}{259} + \frac{A_{40_K}}{4810} \leq 1 \qquad (5)$$

The value of this index must be less than unity in order to keep the radiation hazard insignificant. The maximum value of hazard index equal to unity corresponds to the upper limit of radium equivalent activity (370 (Bq/kg). The values varied from 0.34±0.01 to 0.8±0.013 with mean value of 0.53±0.011 (see table 2). The mean values of hazard index for soil samples were 0.25±0.01 for Jordan [10], 0.38±0.16 for Thailand [11], and 0.13 for Saudi Arabia [13.

Figure 6 shows that the thorium contribution are most significant compared with other radionuclides.

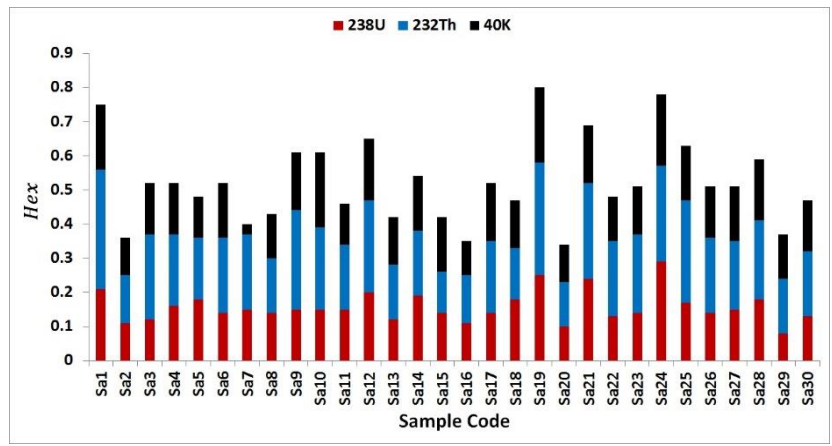

**Figure 6.** The external hazard index for all soil sample.

## 4. Conclusions

The radiological doses as absorbed gamma-ray dose rate, annual effective dose equivalent and annual gonadal dose equivalent resulting from $^{238}U$, $^{232}Th$ and $^{40}K$ families in soil samples from Kerbala city have been evaluated using gamma-ray spectroscopy system using NaI (Tl). To determine the radiological risk gamma radiation representative level index and external hazard index were evaluated. The obtained results show that the average values of dose rate, effective dose and annual gonadal dose are higher than the worldwide average. Since the external hazard index less than unity, therefore no significant radiological hazard for all soil samples in the study area.

**Table 2.** The radiological doses and hazard indices for all soil samples.

| Sample code | D (nGy/h) | AEDE (μSv/y) | AGDE (μSv/y) | $I_{\gamma r}$ | $H_{ex}$ |
|---|---|---|---|---|---|
| Sa1 | 131.02±2.26 | 160.68±2.77 | 921.91±16.03 | 2.07±0.04 | 0.76±0.013 |
| Sa2 | 63.65±1.82 | 78.06±2.23 | 449.88±12.93 | 1±0.03 | 0.37±0.01 |
| Sa3 | 90.36±2.01 | 110.82±2.47 | 639.05±14.25 | 1.43±0.03 | 0. 52±0.012 |
| Sa4 | 88.80±2.00 | 108.9±2.45 | 626.25±14.19 | 1.4±0.03 | 0.51±0.011 |
| Sa5 | 84.69±1.98 | 103.86±2.43 | 594.16±13.99 | 1.32±0.03 | 0.49±0.011 |
| Sa6 | 90.85±2.01 | 111.42±2.47 | 642.73±14.27 | 1.43±0.03 | 0.52±0.012 |
| Sa7 | 66.47±1.87 | 81.52±2.29 | 457.75±13.20 | 1.04±0.03 | 0.4±0.011 |
| Sa8 | 74.18±1.90 | 91.75±2.33 | 528.27±13.50 | 1.17±0.03 | 0.43±0.011 |
| Sa9 | 106.56±2.12 | 130.69±2.60 | 752.83±15.01 | 1.69±0.03 | 0.62±0.012 |
| Sa10 | 106.89±2.10 | 131.09±2.58 | 760.41±14.92 | 1.69±0.03 | 0.61±0.012 |
| Sa11 | 80.26±1.95 | 98.43±2.39 | 564.62±13.81 | 1.26±0.03 | 0.46±0.011 |
| Sa12 | 111.90±2.15 | 137.23±2.64 | 788.16±15.23 | 1.76±0.03 | 0.65±0.012 |
| Sa13 | 74.48±1.89 | 91.34±2.32 | 527.28±13.43 | 1.17±0.03 | 0.43±0.011 |
| Sa14 | 94.82±2.04 | 116.29±2.50 | 667.98±14.43 | 1.48±0.03 | 0.54±0.012 |
| Sa15 | 74.29±1.88 | 91.11±2.31 | 528.34±13.36 | 1.16±0.03 | 0.42±0.011 |
| Sa16 | 60.52±1.80 | 74.22±2.21 | 426.95±12.80 | 0.95±0.03 | 0.35±0.01 |
| Sa17 | 89.42±1.99 | 109.66±2.44 | 633.60±14.15 | 1.41±0.03 | 0.51±0.011 |
| Sa18 | 82.06±1.95 | 100.64±2.39 | 577.89±13.82 | 1.28±0.03 | 0.74±0.011 |
| Sa19 | 138.25±2.31 | 169.55±2.83 | 972.97±16.34 | 2.17±0.04 | 0.8±0.013 |
| Sa20 | 58.89±1.79 | 72.22±2.20 | 416.49±12.70 | 0.93±0.03 | 0.34±0.01 |
| Sa21 | 119.47±2.20 | 146.52±2.70 | 838.32±15.58 | 1.87±0.03 | 0.69±0.013 |
| Sa22 | 81.93±1.96 | 100.48±2.40 | 577.64±13.88 | 1.3±0.03 | 0.48±0.011 |
| Sa23 | 86.70±1.99 | 106.33±2.44 | 611.13±14.12 | 1.37±0.03 | 0.5±0.011 |
| Sa24 | 135.28±2.28 | 165.91±2.80 | 950.80±16.16 | 2.11±0.04 | 0.78±0.013 |
| Sa25 | 107.52±2.13 | 131.86±2.61 | 757.00±15.06 | 1.7±0.03 | 0.63±0.012 |
| Sa26 | 88.06±1.99 | 108±2.44 | 622.90±14.14 | 1.39±0.03 | 0.51±0.011 |
| Sa27 | 90.06±2.00 | 110.45±2.45 | 637.15±14.21 | 1.42±0.03 | 0.52±0.011 |
| Sa28 | 102.69±2.09 | 125.94±2.56 | 725.52±14.80 | 1.62±0.03 | 0.59±0.012 |
| Sa29 | 63.34±1.81 | 77.68±2.22 | 450.58±12.87 | 1.01±0.03 | 0.36±0.01 |
| Sa30 | 81.52±1.94 | 99.98±2.38 | 576.91±13.81 | 1.28±0.03 | 0.47±0.011 |
| Min | 58.89±1.79 | 72.22±2.20 | 416.49±12.70 | 0.93±0.03 | 0.34±0.01 |
| Max | 138.25±2.31 | 169.55±2.83 | 972.97±16.34 | 2.17±0.04 | 0.8±0.013 |
| Mean±S.D. | 90.83±2.00 | 111.89±2.46 | 640.85±15.44 | 1.43±0.031 | 0.53±0.011 |